\documentclass[twocolumn,showpacs,preprintnumbers,amsmath,amssymb]{revtex4}
\usepackage{graphicx}
\usepackage{dcolumn}
\usepackage{bm}
\newcommand{\Ref}[1]{(\ref{#1})}
\newcommand{\Hphys}{\hat{H}_{\textrm{phys}}}
\newcommand{\Hphyscl}{{H}_{\textrm{phys}}}
\newcommand{\sa}{self-adjoint}
\newcommand{\Sas}{Self-adjointness}

\newcommand{\odd}[3]{\frac{d^#3#1}{d#2^#3}}
\newcommand{\nn}{\nonumber \\ }
\newcommand{\paren}[1]{{\left( #1 \right)}}

\newcommand{\wb}{\overline}
\newcommand\sx{{\sqrt |x|}}
\newcommand{\HH}{\mathcal{H}}
\newcommand{\R}{\mathbb{R}}
\newcommand{\ba}[1]{\overline{#1}}

\begin{document}


\title{Big bounce as scattering of wave function at big crunch}

\author{Fumitoshi Amemiya}
 \email{famemiya@rk.phys.keio.ac.jp}
\author{Tatsuhiko Koike}
 \email{koike@phys.keio.ac.jp}
\affiliation{Department of Physics, Keio University, 3-14-1 Hiyoshi, Kohoku-ku, 223-8522 Yokohama, Japan }

\date{\today}

\begin{abstract}
 A gauge-invariant quantum theory of the Friedmann-Robertson-Walker (FRW) universe 
 with dust is studied in terms of the Ashtekar variables.
We use the reduced phase space quantization which has following advantages: 
(i) fundamental variables are all gauge invariant, 
(ii) there exists a physical time evolution of gauge-invariant quantities, 
so that the problem of time is absent and 
(iii) the reduced phase space can be quantized in the same manner as in ordinary quantum mechanics.
In the FRW model, the dynamical components of the Ashtekar variables are given by 
a single quantity $p$ and its conjugate momentum, where $p$
is related to the scale factor $a$ as $a\propto \sqrt{|p|}$
and its sign gives the orientation of triads.
We solve a scattering problem in terms of ingoing and outgoing energy eigenstates.
We show that 
the incident wave is reflected in rate $1/4$ and transmitted in rate $3/4$ 
at the classical singularity $p=0$.
Analyzing the dynamics of a wave packet, we show that the classical initial singularity 
is replaced by a big bounce in quantum theory.
A possible interpretation of the result is that the wave function of the universe
has been in a superposition of states representing right-handed and left-handed 
systems before the big bounce.
\end{abstract}

\maketitle
\section{Introduction}
The initial singularity is inevitable in most of the classical
cosmological models. 
One of the motivations of 
quantum cosmology is to shed light on quantum nature of the singularity.
However, there exists potential problems that have not been completely resolved yet. 
A problem is about what should be interpreted as observables in classical and quantum gravity~\cite{Bergmann, Rovelli}.
A canonical formulation of general relativity (GR) is a constrained system with 
first-class constraints in which the spacetime diffeomorphisms are interpreted as
gauge transformations. 
In gauge theories, only gauge-invariant quantities are observables.
However, there are technical and conceptual difficulties in the realization of the idea
especially in GR (see, e.g.,~\cite{Thiemann book} and references therein).
In many works, gauge-variant quantities are used as observables.
This issue must be seriously considered especially in quantum gravity because it is
substantially related to the problem of time~\cite{pot}. 

Recently, a gauge-invariant quantum theory of the Friedmann-Robertson-Walker
(FRW) universe with dust has been proposed~\cite{fatk}.
There a variable associated with the dust serves as an internal time~\cite{BK}.
The so-called relational formalism~\cite{BD, reduced phase space} has been used
to construct the classical reduced phase space spanned by gauge-invariant
quantities, and then the system has been quantized in the same manner as in
ordinary quantum mechanics. 
In the method, the phase space is reduced at the classical level and then
quantized, so that the procedure is completely different from the Dirac
quantization~\cite{dirac} where one quantizes the unconstrained phase space at
first and then imposes the quantum constraints on physical states.  
Although the full gauge-invariant quantum theory of gravity has not been 
obtained at present, the method gives a possible resolution to
the problem of time and observable at least in the context of minisuperspace
models. 

It has also been shown in~\cite{fatk} that the
initial singularity is replaced by a big bounce. 
In the analyses, the scale factor $a$ itself has been chosen as one of the canonical
variables and its range has been restricted to $a\ge 0$ to avoid the negative scale factor.
One may think that the big bounce is caused by the boundary conditions on wave functions
due to the restriction,
that is, the result depends on the variable one choose.
In fact, since canonical formulations that were equivalent classically yield different quantum theories,
it is quite interesting to investigate the presence of the big bounce by using other
well-motivated variables with unrestricted range.

In this paper, we shall construct and analyze the gauge-invariant quantum theory
of the flat FRW universe with the dust in terms of the Ashtekar variables~\cite{Ashtekar
  formulation, Barbero}.
The Ashtekar formulation has been frequently used in quantum gravity (e.g.~\cite{AshtekarQG})
and, in particular, is employed as a starting point of the quantization 
in loop quantum gravity~\cite{Thiemann book, LQG}.
In the Ashtekar formulation, 
a SU(2) connection $A^i_a$ and a densitized triad $E_i^a$ constitute a canonical pair
where, in the FRW model, the dynamical component of the triad is given by a single 
quantity $p$ such that $a\propto \sqrt{|p|}$.
The range of $p$ is the whole real line, which is doubled from that of $a$,
and the sign of $p$ determines an orientation of the triad. 
We use the reduced phase space quantization method as in the previous work,
which is different from loop quantization.
In the present system, since the Hamiltonian becomes singular at $p=0$,
we discuss carefully the self-adjointness of the Hamiltonian and the space of wave functions.
In view of the quantization,
the Ashtekar variables are shown to be quite natural because
there are no need to impose any boundary conditions on wave functions at the singularity.

We shall then analyze the obtained quantum theory
to reveal the quantum nature of the initial singularity.
First, we solve a scattering problem in terms of ingoing and outgoing energy eigenstates.
The analysis shows that a part of the incident wave is reflected and the rest is transmitted at $p=0$.
Second, we consider the motion of a wave packet and evaluate the expectation value of the scale factor.
It is shown that the expectation value has a non-zero minimum, that is, 
the initial singularity is replaced by a big bounce in quantum theory.
The remarkable point is that the big bounce mixes the states representing 
right-handed and left-handed systems.
One of the obtained scenarios is that if the present
universe is in the right-handed state,
the past universe before the big bounce was in a superposition 
of states representing right-handed and left-handed systems.

The organization of the present paper is as follows.
In Sec.~\ref{sec2}, we introduce the classical reduced phase space of the FRW universe with dust.
In Sec.~\ref{sec3}, we quantize the reduced system and obtain the Schr\"{o}dinger equation.
In Sec.~\ref{sec4}, we analyze the obtained quantum theory to see the dynamics of the universe.
In Sec.~\ref{sec5}, we conclude the paper, and Sec.~\ref{sec6} is devoted to discussion. 
   
In this paper we adopt the unit in which $c=1$.

\section{Classical system \label{sec2}}

In this section, we shall construct the reduced phase space of the classical FRW
cosmology. 

First, we present the classical system of the FRW universe with dust in terms of
the Ashtekar variables. 
Second, we briefly review the relational formalism, 
emphasizing that the relational formalism for a deparametrized system enables
one to explicitly construct 
gauge-invariant quantities and a physical Hamiltonian which generates the time
evolution thereof. 
Third, we construct the reduced phase space of the FRW universe with dust by
using the relational formalism. 

 
\subsection{Classical phase space}      
In the Ashtekar formulation, the variables $(A^i_a, E^a_i)$ form a canonically conjugate pair where $A^i_a$ is a $\textrm{SU}(2)$ connection and $E^a_i$ is an orthonormal triad with density weight $1$.
We here consider the flat FRW universe which is described by the metric
\begin{align}
ds^2=-dt^2+a^2(t)(dx^2+dy^2+dz^2).
\end{align}
In the flat FRW model, the Ashtekar variables can be written in terms of only one independent components $\tilde{c}$ and $\tilde{p}$,
\begin{align}
A^i_a=\tilde{c}(t)\omega^i_a, \quad E_i^a=\tilde{p}(t)X^a_i,
\end{align}
where $\omega^i$ are bases of left invariant one-forms and $X_i$ are invariant vector fields dual to the one-forms.
These variables have relations to the scale factor $a$ such that
\begin{align}
|\tilde{p}|=a^2, \quad \tilde{c}=\textrm{sgn}(p)\frac{\gamma}{N}\dot{a},
\end{align}
where $\gamma$ is the so-called Barbero-Immirzi parameter~\cite{Barbero, Immirzi}, $N$ is the lapse function and the dot denotes the derivative with respect to $t$.
Note that, while the scale factor is restricted to be nonnegative,
$\tilde{p}$ ranges over the entire real line, carrying an orientation of triads
determined by the sign of $\tilde{p}$.

Since the three-space integral, defined as $V:=\int dx^3$, diverges in the flat FRW model with $\mathbb{R}^3$ topology, one must somehow get rid of this divergence.
There are at least two ways to treat this issue.
The first is to consider a compact universe.
The second is to introduce a finite cell and restrict the range of the integral to this cell.
The physical interpretation of the resulting theory becomes clearer in the first case
because one can interpret $|p|^{(1/2)}$ defined below as the length of the shortest nontrivial loops of the universe.
Thus, we take the first option and in particular we only consider the case of three-dimensional torus, where we take a cube of coordinate range $0\le x,y,z\le V^{\frac{1}{3}}$,
and identify the opposite faces.

If we define new variables as
$p:=V^{\frac{2}{3}}\tilde{p}$ and $c:=V^{\frac{1}{3}}\tilde{c}$,
the gravitational action is written as
\begin{align}
S_{\textrm{grav}}=\int dt \left[\frac{3}{\kappa \gamma}p \dot{c} + NH_{\textrm{grav}}\right],
\end{align}
where $\kappa=8\pi G$ and the Hamiltonian constraint $H_{\textrm{grav}}$ takes the form
\begin{align}
H_{\textrm{grav}}=-\frac{3}{\kappa \gamma^2}c^2\sqrt{|p|}.
\end{align}
In terms of the new variables, the Poisson brackets are independent of $V$:
\begin{align}
\left\{c,p\right\}=\frac{\kappa\gamma}{3}.
\end{align}     

Turning to matter couplings, we consider the dust introduced by Brown and Kucha\v{r}~\cite{BK}.
The dust is dynamically coupled to gravity and the action of the dust in the FRW spacetime takes the form
\begin{align}
S_{\textrm{dust}}=\int dt \left[ P_T\dot{T}- NH_{\textrm{dust}}   \right].
\end{align}
Here, $T$ is the proper time measured along the particle flow lines when the equations of motion hold, $P_T$ is its conjugate momentum and 
\begin{align}
H_{\textrm{dust}}=P_T.
\end{align}
In particular, it can be shown by solving the equations of motion that $T$ coincides with the cosmological proper time in isotropic and homogeneous models.
It serves as a natural time variable in the construction of the reduced phase space.

Combining these results, we now have the total action for gravity plus the dust,
\begin{align}
S_{\textrm{tot}}=\int dt \left[ \frac{3}{\kappa \gamma} p\dot{c}+P_T\dot{T}-NH_{\textrm{tot}} \right],
\end{align}
where
\begin{align}
H_{\textrm{tot}}=H_{\textrm{grav}}+H_{\textrm{dust}}
=-\frac{3}{\kappa \gamma^2}c^2\sqrt{|p|}+P_T=0.\label{constraint}
\end{align}
The form of the constraint (\ref{constraint}) is called deparametrized form, 
which has a special feature as will be seen in the discussion below.


\subsection{Relational formalism}
We here give a brief review of the relational formalism,
which is a method to construct gauge-invariant quantities in constrained systems
(see~\cite{BD, reduced phase space} for details).
From now on, we assume that the system has only one constraint for simplicity.

The key observation of the relational formalism to define gauge-invariant
quantities is as follows.  
Take two gauge-variant functions $F$ and $T$ on the phase space, and choose one of the functions $T$ as a clock.
Then, the value of $F$ at $T=\tau$ is gauge-invariant even if $F$ and $T$
themselves are gauge variant. 
That is, a selected function $T$ serves as a clock and the relation between $T$ and other variables is interpreted as time evolution.
 
Let us now move to the mathematical definition of the gauge-invariant quantities in the relational formalism. 
Suppose a phase space has a 2$n$-dimension ($n \ge 2$), and there are canonical coordinates $(q^a, p_a, a=1,\cdots ,n)$ such that
$\{q^a, p_b\}=\delta^a_b$.
We will denote a first-class constraint by $H$ and a phase space point by $y=(q^a,p_a)$.
Under the gauge transformation generated by $H$, a point $y$ is mapped to 
$y \mapsto \alpha_H^t (y)$, where $t$ is a gauge parameter.
That is, $\alpha_H^t(y)$ is a gauge flow generated by $H$ starting from $y$.

\begin{figure}
\includegraphics[width=65mm]{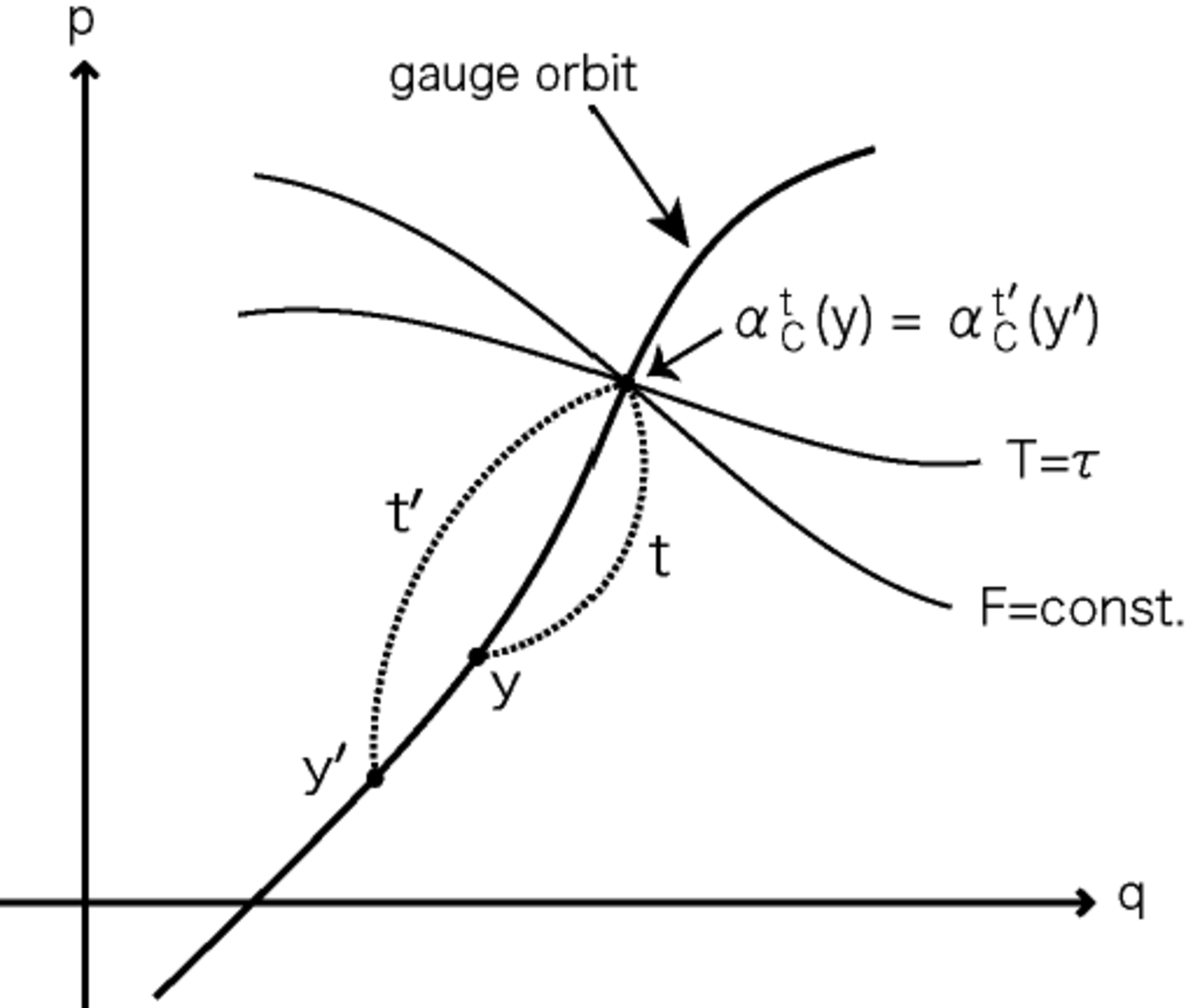}
\caption{\label{alpha}The action of the map $\alpha_H^t$ on phase space points and 
the gauge-invariance of $O_{F}^{\tau}(y)$. }
\end{figure}

Then we can define the gauge-invariant quantity $O_F^\tau(y)$ as
\begin{align}
O_{F}^\tau(y) := F(\alpha_H^t(y)) | _{T(\alpha_H^t(y))=\tau}.\label{gi}
\end{align}
Here, $F(\alpha_H^t(y))$ can be written as a series $F(\alpha_H^t(y)) = \sum_{n=0}^{\infty}\frac{t^n}{n!}\{H,F\}_{(n)}(y)$,
where $\{H,F\}_{(0)}:=F$ and $\{H,F\}_{(n+1)}:=\left\{H,\{H,F\}_{(n)} \right\}$. 
The definition (\ref{gi}) gives a manifestly gauge-invariant quantity because
$O_F^\tau(y)$ is constant on each gauge orbit.
Indeed, as illustrated in Fig.\ref{alpha}, if $y$ and $y^{\prime}$ are on the same gauge orbit, there is $t^{\prime}$ such that $\alpha_{C}^{t}(y)=\alpha_{C}^{t^{\prime}}(y^{\prime})$ and 
\begin{align}
O_F^{\tau}(y)&=F\left(\alpha_C^t(y)\right)\bigr|_{T\left(\alpha_C^t(y)\right)=\tau}\nonumber\\
&=F\left(\alpha_C^{t^{\prime}}(y^{\prime})\right)\Bigr|_{T\left(\alpha_C^{t^{\prime}}(y^{\prime})\right)=\tau}=O_F^{\tau}(y^{\prime}).
\end{align}

A constraint equation $H=0$ is said to be of deparametrized form if it is written as 
\begin{align}
H(q^a,T,p_a,P_T)=P+h(q^a,p_a)=0
\end{align}
with some phase space coordinates $\{q^a,T;p_a,P_T\}$. 
In the deparametrized theories, the reduced phase space is spanned by the gauge-invariant quantities $\left(O_{q^a}^\tau(y),O_{p_a}^\tau(y)\right)$ associated with $q^a$ and $p_a$ with the simple symplectic structure
\begin{align}
\left\{O_{q^a}^\tau(y),O_{p_b}^\tau(y)\right\}=\delta^a_b.\label{poisson}
\end{align}
The physical Hamiltonian $H$ is obtained by replacing $q^a$ and $p_a$ in $h(q^a,p_a)$ with $O_{q^a}^\tau(y)$ and $O_{p_a}^\tau(y)$,
\begin{align}
H\left(O_{q^a}^\tau(y),O_{p_a}^\tau(y)\right):=h\left(O_{q^a}^\tau(y),O_{p_a}^\tau(y)\right).\label{H}
\end{align}
The Hamiltonian generates the time evolution of the gauge-invariant quantity associated with
a function $F$ which depends only on $q^a$ and $p_a$:
\begin{align}
\frac{\partial O_{F}^\tau(y)}{\partial \tau}= \left\{H,O_{F}^\tau(y)\right\}.\label{hamilton}
\end{align}

As we have seen, the relational formalism enables one to construct gauge-invariant quantities 
and extract their time evolution in the deparametrized case.
Therefore, the method provides us a possible resolution to the problem of time and observable.


\subsection{Reduced phase space of Friedmann-Robertson-Walker universe with dust}
Let us now construct the reduced phase space of the FRW universe with dust.
As mentioned above, 
if the constraint is written in the deparametrized form,
one can construct the reduced phase space coordinatized by gauge-invariant quantities and extract their time evolution by using the relational formalism.

In the present case, it is natural to choose the function $T$ as the clock variable.
Then, the reduced phase space is coordinatized by the gauge-invariant quantities $C(\tau):=O_c^{\tau}(y)$ and $P(\tau):=O_p^{\tau}(y)$ associated with $c$ and $p$ with very simple symplectic structure  
\begin{align}
\left\{C(\tau),P(\tau)\right\}=\frac{\kappa\gamma}{3}.\label{poisson}
\end{align}
Moreover, we can obtain the physical Hamiltonian $H_{\textrm{phys}}$ 
by replacing $c$ and $p$ in $H_{\textrm{grav}}(c,p)$ with $C$ and $P$,
\begin{align}
H_{\textrm{phys}}=-\frac{3}{\kappa \gamma^2}C(\tau)^2\sqrt{|P(\tau)|}.\label{ham}
\end{align}
The Hamiltonian (\ref{ham}) generates time evolution of a gauge invariant quantity $F$ as
\begin{align}
\frac{dF}{d\tau}=\left\{ H_{\textrm{phys}},F \right\}; \ F=F\left(C(\tau),P(\tau)\right).
\end{align}

So far, we have constructed the reduced phase space of the FRW universe with dust.
We will consider the quantization of the system in the next section.

\section{quantization \label{sec3}}
In this section, we shall quantize the system on the reduced phase space obtained in
the previous section. 

The quantization procedure is straightforward as in the ordinary quantum mechanics 
because we need not to deal with the constraint.
Now the physical variables are operators 
and 
the Poisson bracket $\{\bullet,\bullet\}$ is replaced with 
the commutation relation $(1/i\hbar)[\bullet,\bullet]$. 
Thus (\ref{poisson}) becomes the canonical commutation relation 
\begin{align}
[\hat{C},\hat{P}]=\frac{i\kappa\gamma\hbar}{3}.
\end{align}
The evolution of the state $\Psi$ of the universe 
is described by the Schr\"odinger equation
\begin{align}
  i\hbar\frac{\partial\Psi}{\partial \tau}=\hat{H}_{\textrm{phys}}\Psi, 
  \label{sch0}
\end{align}
where $\Hphys:=\Hphyscl(\hat C,\hat P)$. 

Although the prescription above formally 
gives a quantum theory of the universe, 
the theory is not well defined or uniquely determined until we specify the
operator 
ordering, the Hilbert space and the 
the domain of the operator $\Hphys$ 
which makes $\Hphys$ {\sa}. 
{\Sas}, beyond Hermiticity, of $\Hphys$ is 
substantial 
in quantum mechanics because it is necessary for unique development of the
quantum state and the conservation of probability. 

Let us choose the ordinary Schr\"{o}dinger representation in which the
operators $\hat{P}$ and $\hat{C}$, respectively, 
act on a wave function $\Psi(P)$ in the following way: 
\begin{align}
\hat{P}\Psi(P)=P\Psi(P),\quad \hat{C}\Psi(P)=\frac{i\hbar\kappa\gamma}{3}\frac{\partial \Psi(P)}{\partial P}.
\end{align}

As a concrete example, we choose the following operator ordering for the
Hamiltonian, 
\begin{align}
\hat{H}_{\textrm{phys}}=-\frac{3}{\kappa \gamma^2}\sqrt{|\hat{P}|}\hat{C}^2. 
\label{eq-ord1}
\end{align}
Then the Schr\"{o}dinger equation \Ref{sch0} takes the simplest form, 
\begin{align}
i\hbar\frac{\partial \Psi}{\partial
  \tau}=
\frac{\kappa\hbar^2}{3}\sqrt{|P|}\frac{\partial^2 \Psi}{\partial P^2}.
\label{sch1}
\end{align}
Since the present Hamiltonian is different from the ordinary kinematical term,
we choose the Hilbert space as 
$\mathcal{H}=L^{2}( \mathbb{R},|P|^{-\frac{1}{2}}dP)$
the set of square integrable functions of $P$ with respect to the 
measure $|P|^{-\frac{1}{2}}dP$
in order to make the Hamiltonian \Ref{eq-ord1} Hermitian up to surface term. 
We shall rewrite the Schr\"{o}dinger equation 
\Ref{sch1} in dimensionless variables 
$x:=P/\left( \frac{3}{8\pi} \right)^{\frac{2}{3}}l_{\mathrm{Pl}}^2$ and
$\tilde{\tau}:=\tau/t_{\mathrm{Pl}}$, 
\begin{align}
i\frac{\partial \Psi(x,\tilde{\tau})}{\partial \tilde{\tau}}=\sqrt{|x|}\frac{\partial^2
  \Psi(x,\tilde{\tau})}{\partial x^2},\label{dimensionless} 
\end{align}
where $l_{\mathrm{Pl}}$ and $t_{\mathrm{Pl}}$ are 
the Planck length and the Planck time, respectively. 

Though the Hamiltonian 
$\Hphys$ is somewhat singular at the origin $x=0$, 
it is indeed {\sa} in $\mathcal{H}$, 
which we shall show in Appendix~\ref{sa}.
Note that 
we do not need any special boundary conditions for the 
wave function $\Psi$ at $x=0$.

\section{dynamics of the universe\label{sec4}}
In this section, we shall investigate the obtained one-dimensional quantum system
to analyze the quantum nature of the initial singularity.
First, we solve a scattering problem in terms of ingoing and outgoing energy eigenstates.
Second, we analyze the motion of a wave packet and evaluate the expectation value of the scale factor.
Third, we construct a special analytic wave function and investigate its time evolution.

\subsection{Reflection and transmission rates}

We here calculate the reflection and transmission rates
at the origin $x=0$.
From the Schr\"{o}dinger equation (\ref{dimensionless}), the energy eigenvalue equation reads
\begin{align}
\Hphys \psi(x)=\sqrt{|x|}\frac{d^2 \psi(x)}{d x^2}=E\psi(x).\label{e-eq1}
\end{align} 
We here note that $\sqrt{|x|}d^2/dx^2$ is a negative definite self-adjoint operator 
on $L^2(\mathbb{R},|x|^{-\frac{1}{2}}dx)$.
The negative definiteness, $\langle \psi |\Hphys|\psi\rangle\le 0$ for all $\psi$, 
can be easily shown by integration by parts,
and it is consistent with the fact that the gravitational energy never takes positive values
in classical theory.
Then, we can rewrite the energy eigenvalue equation (\ref{e-eq1}) as
\begin{align}
\sqrt{|x|}\frac{d^2 \psi(x)}{d x^2}=-|E|\psi(x).\label{e-eq2}
\end{align} 

To obtain the eigenfunctions, we make use of the fact that
the equation (\ref{e-eq2}) can be transformed
into Bessel's differential equation.
Let us first find the solution of Eq.~(\ref{e-eq2}) for $x>0$.
In terms of $\phi=\psi/\sqrt{x}$ and $z=\frac{4}{3}\sqrt{|E|}x^{\frac{3}{4}}$, 
Eq.~(\ref{e-eq2}) yields
\begin{align}
z^2 \frac{d^2 \phi}{dz^2}+z\frac{d\phi}{dz}+(z^2 - \frac{4}{9})\phi=0.
\end{align}
The solution is written as
\begin{align}
\phi(z)&=A_1 J_{2/3}(z) + A_2
 J_{-2/3}(z)\\
    &=B_1 H^{(1)}_{2/3}(z) + B_2
     H^{(2)}_{2/3}(z),
\end{align}
where $A_1$, $A_2$, $B_1$ and $B_2$ are arbitrary constants, 
$J_{\pm 2/3}$ are the Bessel functions
and $H^{(1,2)}_{2/3}$ are the Hankel functions.
Thus, the eigenfunctions for $x>0$ take the form
\begin{align}
\psi(x)
&=\sqrt{x}\left[A_1 J_{2/3}\left(\frac{4}{3}\sqrt{|E|}x^{\frac{3}{4}}\right) 
+ A_2  J_{-2/3}\left(\frac{4}{3}\sqrt{|E|}x^{\frac{3}{4}}\right)\right]\\
&=\sqrt{x}\left[B_1 H^{(1)}_{2/3}\left(\frac{4}{3}\sqrt{|E|}x^{\frac{3}{4}}\right) 
+ B_2 H^{(2)}_{2/3}\left(\frac{4}{3}\sqrt{|E|}x^{\frac{3}{4}}\right)\right].
\end{align}
For later use, we here explicitly give the definition of the special functions; 
$J_\nu(z):=\sum_{n=0}^{\infty}\frac{(-1)^n}{n!\Gamma(n+\nu+1)}\left(\frac{z}{2}\right)^{2n+\nu}$
and $H_\nu^{(1,2)}(z):=J_\nu(z)\pm i N_{\nu}(z)$ where
the Neumann function $N_{\nu}(z)$ is defined as 
$N_{\nu}(z):=\frac{J_\nu(z)\cos(\nu\pi)-J_{-\nu}(z)}{\sin(\nu\pi)}$.
To obtain the energy eigenfunctions in the whole $x$, we connect the functions for
$x>0$ and $x<0$ at the origin with the conditions $\psi(-0)=\psi(+0)$ and 
$d\psi/dx(-0)=d\psi/dx(+0)$.
We have
\begin{align}
\psi(x)
=\sqrt{|x|}\bigg[A_1 \mathrm{sgn}(x) J_{2/3}\left(\frac{4}{3}\sqrt{|E|}|x|^{\frac{3}{4}}\right) \nonumber\\
+ A_2  J_{-2/3}\left(\frac{4}{3}\sqrt{|E|}|x|^{\frac{3}{4}}\right)\bigg],\label{eigenstate}
\end{align}
where the first term is an odd function and the second is an even function. 
Here we should note that these energy eigenvalue functions do not lie 
in the Hilbert space $L^2(\mathbb{R},|x|^{-\frac{1}{2}}dx)$. 
This situation is similar to the system of a free 
particle in quantum mechanics and one can expand a physical 
state in terms of the energy eigenstates.

\begin{figure}
\includegraphics[width=65mm]{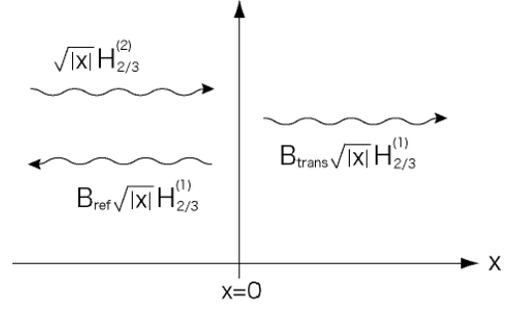}
\caption{\label{fig:stationary} 
The conception of the incident, reflected and transmitted waves. }
\end{figure}

Let us now move to the calculation of the reflection and transmission rates.
We recall here that the Hankel functions have the asymptotic forms 
$H^{(1,2)}_{\nu}(z) \sim \sqrt{\frac{2}{\pi z}}e^{\pm i[z-(\nu+\frac{1}{2})\frac{\pi}{2}]}$
for large $z$.
Thus they are the analogues 
of the exponential functions $e^{\pm i x}$, $z=kx$, in the free particle case.
We consider the following stationary state $\psi(x)$;
(i) the incident wave propagates in the positive $x$ direction 
and its amplitude is set to one, 
and (ii) there are no waves that propagate in the negative $x$ direction for $x\to \infty$,
that is, 
\begin{align}
\psi(x)=&B_{\mathrm{ref}} \sqrt{|x|}H^{(1)}_{2/3}\left(\frac{4}{3}\sqrt{|E|}|x|^{\frac{3}{4}}\right) \nonumber\\
&+  \sqrt{|x|}H^{(2)}_{2/3}\left(\frac{4}{3}\sqrt{|E|}|x|^{\frac{3}{4}}\right)\ \mathrm{for} \  x<0,\\
\psi(x)=&B_{\mathrm{trans}} \sqrt{|x|}H^{(1)}_{2/3}\left(\frac{4}{3}\sqrt{|E|}|x|^{\frac{3}{4}}\right) 
\ \mathrm{for} \  x>0,
\end{align}
where $B_{\mathrm{ref}}$ and $B_{\mathrm{trans}}$ are the amplitudes of the reflected and transmitted waves, 
respectively (See Fig.~\ref{fig:stationary}).
We can calculate the amplitudes by imposing the conditions $\psi(-0)=\psi(+0)$ and
$d\psi/dx(-0)=d\psi/dx(+0)$.
These conditions yield
\begin{align}
B_{\mathrm{ref}}=\frac{i}{i-\sqrt{3}},\quad B_{\mathrm{trans}}=\frac{\sqrt{3}}{i-\sqrt{3}}.
\end{align}
The difference between the phases of the reflected and the transmitted waves
is $-\pi /2$ because $B_{\mathrm{trans}}/B_{\mathrm{ref}}=\sqrt{3}e^{-i\frac{\pi}{2}}$.

In the present system, the conservation law for probability density
$\frac{\partial}{\partial \tilde{\tau}}\rho(x,\tilde{\tau})+\frac{\partial}{\partial x}j(x,\tilde{\tau})=0$
holds where the energy density $\rho(x,\tilde{\tau})$ and its flow $j(x,\tilde{\tau})$
are defined as 
\begin{align}
\rho(x,\tilde{\tau})&:=\frac{|\psi(x,\tilde{\tau})|^2}{\sqrt{|x|}},\\
j(x,\tilde{\tau})&:=i\left[\psi^{\ast}(x,\tilde{\tau})\frac{\partial \psi(x,\tilde{\tau})}{\partial x}-\frac{\partial \psi^{\ast}(x,\tilde{\tau})}{\partial x}\psi(x,\tilde{\tau})\right].
\end{align}
Then, one can define the reflection rate $R$ and the transmission rate $T$ as
\begin{align}
R=\bigg|\frac{j_R(x)}{j_I(x)}\bigg|, \quad T=\bigg|\frac{j_T(x)}{j_I(x)}\bigg|,
\end{align}
where $j_I$, $j_R$ and $j_T$, respectively, are the flow of the probability density 
for the incident wave, the reflected wave and the transmitted wave.
We can obtain the values of $R$ and $T$ at the origin as follows:
\begin{align}
R|_{x=0}&=\bigg|\frac{j_R(0)}{j_I(0)}\bigg|=|B_{\mathrm{ref}}|^2=\frac{1}{4},\\
T|_{x=0}&=\bigg|\frac{j_T(0)}{j_I(0)}\bigg|=|B_{\mathrm{trans}}|^2=\frac{3}{4}.
\end{align}
Note that $R|_{x=0}$ and $T|_{x=0}$ do not depend on the energy eigenvalue $E$.
So far, it has been shown that the incident wave reflects in rate $1/4$ and 
transmits in rate $3/4$.
That is, although we do not impose any boundary conditions at the origin,
neither a complete reflection nor a complete transmission can occur in the present model.

\subsection{Time evolution of wave packet and scale factor}

Let us now analyze the dynamics of a wave packet.
The procedure is as follows.
First, we prepare an initial wave packet $\Psi(x,0)$ at some nonzero $x$.
Then, we numerically evolve it backward in time by the Schr\"{o}dinger equation (\ref{dimensionless})
and evaluate the expectation value of $|x|$ as a function of the internal time $\tilde{\tau}$.
Here we consider $|x|$ because both the positive and negative $x$
correspond to the universe of the same size with different orientation of triads.

The numerical methods used here are the forth-order Runge-Kutta method in the time integration 
and the midpoint difference method for the spatial differentiation.
However, we need a special care for several points near the origin 
because of the singular nature of the Hamiltonian.
The midpoint difference method does not give a good approximation at the origin,
because there the right hand side of Eq.~(\ref{dimensionless}) incorrectly becomes 
zero all the time.
Thus, we use the least square method to fit the wave function near the origin.
We consider the eigenfunction expansion of the wave function 
and take the first five terms in the series as the fitting function $f(x)$:
\begin{align}
f(x)=c_1 + c_2 x + c_3 |x|^{\frac{3}{2}} + c_4 \mathrm{sgn}(x) |x|^{\frac{5}{2}} +c_5 |x|^3.
\end{align}
Then the right hand side of Eq.~(\ref{dimensionless}) can be evaluated as
\begin{align}
\sqrt{|x|}\frac{d^2 f}{dx^2}=\frac{3}{4}c_3+\frac{15}{4}c_4x+6c_5|x|^{\frac{3}{2}}.
\end{align}

\begin{figure}
\includegraphics[width=80mm]{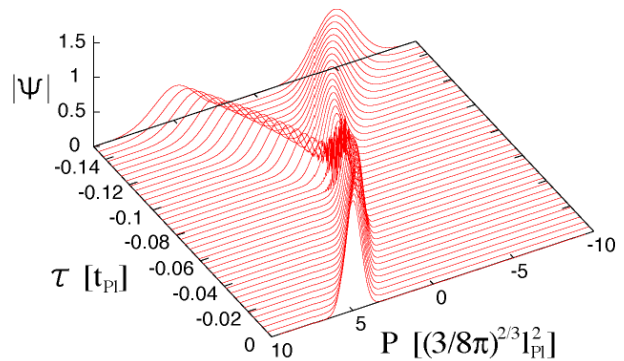}
\caption{\label{fig:wave}
The absolute value of the wave function is plotted as a function of $\tau$ and $P$. }
\end{figure}

\begin{figure}
\includegraphics[width=80mm]{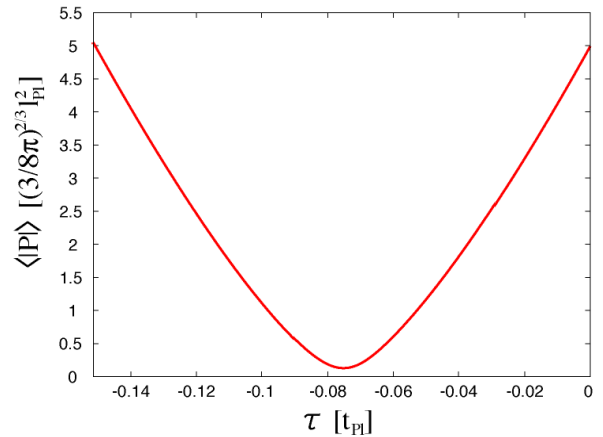}
\caption{\label{fig:exp}  
The expectation values of $P$ is plotted as a functions of the time $\tau$.}
\end{figure}

For simplicity, we here choose the initial wave function as a Gaussian wave packet 
\begin{align} 
\Psi(x,0)=C_0 \exp\left(-\frac{(x-x_0)^2}{2\sigma^2}-ik_0x\right),\label{init_wave}
\end{align}
where $C_0$ is the normalization constant.
Although the other choices are possible, one can obtain qualitatively same results
irrespective of the choice.
In the numerical calculation below, we set the initial values as 
$x_0=5$, $k_0=20$ and $\sigma=0.5$.
Fig.~\ref{fig:wave} shows the absolute value of the wave function as a function of $P$ and $\tau$,
and the expectation value of $|P|$ is plotted as a function of the time $\tau$ in Fig~\ref{fig:exp}.

We can see from Fig.~\ref{fig:wave} that, as is expected from the previous discussion,
a part of the wave packet is reflected and
the rest part is transmitted at the origin.
We here remind that the sign of $x$ determines an orientation of triads, 
which correspond to a right-handed and left-handed systems respectively.
Thus, the result indicates that if the present state of the universe is in a right-handed system,
the past state is in superposition of the states of a right-handed and left-handed
systems.

This calculation allows another interpretation
because $\Psi^{\ast}(x,-\tilde{\tau})$ is the time-reversal solution of the Schr\"{o}dinger equation.
If we take complex conjugate of the initial state (\ref{init_wave}) and set 
$\tilde{\tau}=-\tilde{\tau}^{\prime}$,
we can consider the initial 
wave packet as a state in the past and the above calculation as forward evolution in time.
That is, if a past state of the universe has been in a right-handed system,
the present state is in superposition of the states of a right-handed and 
left-handed systems.

As for the expectation value of $|P|$,
Fig.~\ref{fig:exp} indicates that the expectation value never goes to zero and bounces at 
a nonzero minimum.
That is, the initial singularity is replaced by a big bounce in the present model.
The result is consistent with the previous work~\cite{fatk}, 
where one does not need any boundary conditions at the origin.
The difference is
that the wave function in this model is a superposition of the states representing the
right-handed and left-handed systems.

\subsection{Special analytic wave function}

Let us construct a special analytic wave function.
We superpose the odd functions in the energy eigenstate (\ref{eigenstate}):
\begin{align}
\psi^{\mathrm{odd}}_E(x)
=\mathrm{sgn}(x)\sqrt{|x|} J_{2/3}\left(\frac{4}{3}\sqrt{|E|}|x|\right).
\end{align}
The superposition of the states is written in the form    
\begin{align}
&\Psi(x,\tilde{\tau})=\int^{0}_{-\infty}C(E)e^{-iE\tilde{\tau}}\psi^{\mathrm{odd}}_EdE\nonumber\\
&=\frac{9}{8}\mathrm{sgn}(x)\sqrt{|x|}\int^{\infty}_{0}C(\epsilon)e^{i\frac{16}{9}
\epsilon^2 \tilde{\tau}}\epsilon J_{2/3}\left( |x|^{\frac{3}{4}}\epsilon \right)d\epsilon,
\end{align}
where $\epsilon:=\frac{4}{3}\sqrt{|E|}$ and $C$ is a superposition coefficient.
We choose the function $C(\epsilon)$ to be 
$C(\epsilon)=\frac{8}{9}\epsilon^{2/3} e^{-\alpha \epsilon^2}$
where $\alpha$ is an arbitrary positive constant. 
Then we can use the formula  
$\int^{\infty}_{0}e^{-ax^2}x^{\nu +1}J_{\nu}(bx)dx=\frac{b^{\nu}}{(2a)^{\nu+1}}e^{-\frac{b^2}{4a}}$
where $\textrm{Re}(a)>0$ and $\textrm{Re}(\nu)>-1$ (e.g., \cite{formula}).
The integration yields the wave function
\begin{align}
\Psi(x,\tilde{\tau})= \frac{x}{(2\beta(\tilde{\tau}))^{\frac{5}{3}}}
\exp\left(-\frac{|x|^{\frac{3}{2}}}{4\beta(\tilde{\tau})}\right),\label{odd wave}
\end{align}
where $\beta(\tilde{\tau}):=\alpha - i\frac{16\pi}{9}\tilde{\tau}$.
The wave function (\ref{odd wave}) has double peaks,
which represents a equal-weight superposition of the 
right-handed and left-handed states.

The expectation value of $|x|$ also can be calculated as
\begin{align}
\langle |x| \rangle= \frac{\langle \Psi | |x| |\Psi\rangle}{\langle \Psi |\Psi\rangle}
=\frac{\Gamma \left(\frac{7}{3}\right)}{\Gamma\left(\frac{5}{3}\right)}
\left(\frac{2}{\alpha}\right)^{\frac{2}{3}}\left[ \alpha^2 + \left( \frac{16}{9}\tilde{\tau} \right)^2 \right]^{\frac{2}{3}}.
\end{align}
The solution again shows that the initial singularity is avoided.
For large $\tilde{\tau}$,
the scale factor $a$ is evaluated as
\begin{align}
a \propto \sqrt{\langle |x| \rangle} \propto \tilde{\tau}^{\frac{2}{3}}.
\end{align}
Thus, they are in good agreement with the classical trajectories when the universe is sufficiently large.

\section{conclusions \label{sec5}}
A gauge-invariant quantum theory of the flat FRW universe with dust 
has been studied in terms of the Ashtekar variables.
In the FRW model, the dynamical component of the Ashtekar variables are given by 
a single quantity $p$ and its conjugate momentum $c$, where $p$
is related to the scale factor $a$ as $a\propto \sqrt{|p|}$,
and the sign of $p$ determines an orientation of the triad. 
We have first constructed the classical reduced phase space of the system by using 
the relational formalism and then have quantized the reduced system.
The advantages of the quantization method are as follows: 
(i) fundamental variables $P$ and $C$ are gauge-invariant quantities
associated with the Ashtekar variables $p$ and $c$, 
(ii) a natural time evolution of the gauge-invariant quantities exists, 
so that the problem of time and observable is resolved at least in the deparametrized theory 
and (iii) the reduced phase space can be quantized in the same manner as in 
ordinary quantum mechanics because there are no constraints in the reduced phase space.
In the quantization, we have rigorously discussed and have shown the self-adjointness 
of the Hamiltonian 
$\Hphys=-\frac{3}{\kappa \gamma^2}\sqrt{|\hat{P}|}\hat{C}^2$ which is singular at $P=0$.
There we need not to impose any boundary conditions on wave functions at the singularity
because the range of $P$ is not restricted to be finite as opposed to the scale factor.

In the obtained quantum theory, we have analyzed the quantum nature of the initial singularity.
We have first considered a scattering problem by using the Hankel functions.
We have shown that the incident wave is reflected in rate 1/4 
and transmitted in rate 3/4 at the origin $P=0$ which was the initial singularity in classical theory.
Then, we have investigated the motion of a wave function $\Psi$ of the universe.
In the present paper, we have considered different two scenarios in which the initial 
wave functions $\Psi(P,0)$ have been chosen respectively as a wave packet peaked at some positive $P$ 
and an anti-symmetric wave which has double peaks.
In both cases, it is shown that the expectation value of $P$ has a non-zero minimum,
that is, the initial singularity is replaced by a big bounce in quantum theory.
The interpretation for the case of a wave packet is twofold depending on backward or forward evolution.
From the backward evolution, it follows that
if the present state of the universe is in a right-handed system,
the past state has been in a superposition of the states of a right-handed and left-handed systems, 
and from the forward evolution, it follows that if a past state of the universe has 
been in a right-handed system,
the present state is in a superposition of the states of right-handed and left-handed systems.
In the case of the anti-symmetric wave function, 
the universe has remained in a equal-weight superposition of the 
right-handed and left-handed states all the time.

\section{Discussion \label{sec6}}
The solution scattered 
at the origin $P=0$ may be related to the existing imbalance 
of the parity of the universe. 

A possibility is that the universe after a big bounce is a
superposition of the right-handed and left-handed states,
and one of them is chosen by destruction of the superposition. 
If one takes into account the degrees of freedom of local gravitational field and other fields,
the superposition may be considered as a quantum cat state.
The state is vulnerable to gobservation", 
or other small perturbations, which breaks symmetry of parity. 

Another possibility is that the scattering solutions presented in this
paper correspond to local phenomena in the universe, 
namely, collapses and formation of black holes. 
For example, if one considers situation of 
a Oppenheimer-Snyder collapse where 
the inner spacetime is described by a FRW spacetime, 
one expects that inside the black hole there is a ``big bounce'' 
beyond which there is a universe of the opposite parity. 
Then there may be regions in the universe 
where there are elementary particles 
with opposite parity from those that we presently know.

\section*{ACKNOWLEDGEMENTS}
We would like to thank Professor Seiji Sakoda for fruitful discussions on
singular Hamiltonians. 
One of the authors (FA) is grateful to Mr. Ryosuke Yoshii for 
useful discussions and comments.
This work was supported in part by Global COE Program ``High-Level Global Cooperation
for Leading-Edge Platform on Access Spaces (C12)".

\appendix

\section{Self-adjointness of $\Hphys$}
\label{sa}
Let us show that $\Hphys=\sqrt{|x|}\odd {}x2$ is {\sa} in 
the Hilbert space 
$\HH=L^{2}(\mathbb{R},|x|^{-\frac{1}{2}}dx)$ 
where the inner product is given by 
$(\Phi,\Psi) 
:=\int_{\R} {dx} \ba\Phi\Psi/{\sx}$. 
Difference from the case of 
standard quantum mechanical Hamiltonian 
$H=-\odd{}x2+V(x)$ on $L^{2}(\mathbb{R})$ 
with well-behaved $V(x)$
(e.g. \cite{RS}) 
is 
that we need some care on the point $x=0$. 
However, it turns out to cause no harm. 

We define the domain of $\Hphys$ by 
\begin{align}
  D(\Hphys)=
  \{&
  \Psi\in\HH| 
  \Psi\in C^1(\R), 
  \Psi' \text{ is locally }
  \nn
  &\text{absolutely continuous}, 
  \Hphys \Psi\in\HH\}, 
  \label{eq-domain}
\end{align}
where a prime denotes the derivative, 
and the second differentiation in $\Hphys$ 
is in the sense of local absolute continuity 
(any absolute continuous function is an integral of 
some $L^1$ function). 

Symmetricity of $\Hphys$ is simple. 
One can easily show 
by partial integration
that $\Hphys$ is Hermitian, 
i.e., $(\Phi,\Hphys \Psi)=(\Hphys\Phi,\Psi)$, 
for $\Psi,\Phi\in D(\Hphys)$. 
Then $\Hphys$ is symmetric because $D(\Hphys)$ is dense in $\HH$. 

Let us find the expression and the domain of $\Hphys^\dagger$. 
Let $A$ be the restriction of 
$\Hphys$ on 
$D(A):=\{\Psi\in\HH|\Psi\in C_0^{\infty}(\R)\}$, 
where $C_0^{\infty}(\R)$ is the set of 
smooth functions on $\R$ with compact support. 
For $\Psi\in D(A)$ and 
$\Phi\in D(A^\dagger)$, 
we have
\begin{align}
  \int \frac{dx}{\sx} \paren{\wb{A^\dagger\Phi}}{\Psi}
  =(A^\dagger\Phi,\Psi) 
  =(\phi,A\psi) 
  =\int dx \ba\Phi\Psi''.
  \label{eq-app-1}
\end{align}
Eq. \Ref{eq-app-1} implies 
that the weak second derivative 
$D^2\Phi$ of $\Phi$ is given by 
$
D^2\Phi
={{A^\dagger\Phi}}/{\sx}
$. 
Since 
${A^\dagger\Phi}\in\HH$ and 
$f\in\HH$ implies $f/|x|^{1/4}\in L^2(\R,dx)$, 
we have 
$
D^2\Phi=g/{|x|^{1/4}}$ with 
$g\in{L^2(\R,dx)}$. 

We can show that 
$D^2\Psi$ is locally $L^\alpha(\R,dx)$ with 
$1<\alpha<4/3$. 
By H\"older's inequality, 
we have, for norms on any compact set $K$, 
$
\left\| g/{|x|^{1/4}}\right\|_\alpha^\alpha
=
\left\| (g/{|x|^{1/4}})^\alpha\right\|_1
\leqslant 
\left\|g^\alpha\right\|_{2/\alpha} \||x|^{-\alpha/4}\|_{q}
=
\left\|g\right\|_{2}^{\alpha} \||x|^{-q\alpha/4}\|_{1}^{1/q}
$ 
where 
$q={2}/({2-\alpha})$. 
Since 
${q\alpha}/4<1$, 
$
\||x|^{-q\alpha/4}\|_{1}
$
hence 
$
\left\|g/{|x|^{1/4}}\right\|_\alpha 
$
is finite. 
Thus $D^2\Psi$ is locally $L^\alpha(\R,dx)$. 
Therefore $\Phi$ is locally $W^{2,\alpha}(\R,dx)$. 

By Sobolev's embedding theorem, 
this implies 
$\Phi$ is 
in $C^{1,2-1/\alpha}(\R)$. 
In particular, $\Phi$ is in $C^1(\R)$, i.e., 
$\Phi$ is differentiable everywhere and has a continuous 
derivative $\Phi'$. 
This $\Phi'$ must be locally $W^{1,\alpha}(\R,dx)$ 
because $\Phi$ was locally $W^{2,\alpha}(\R,dx)$. 
Since $\alpha>1$, 
$\Phi'$ is also locally $W^{1,1}(\R,dx)$, 
namely, $\Phi'$ is locally absolutely continuous. 
Thus any $\Phi\in D(A)$ is differentiable and has a locally absolutely
continuous $\Phi'$. 

It follows that any $\Phi\in D(\Hphys^\dagger)$ 
has a locally absolutely continuous derivative $\Phi'$, 
for we have $D(A^\dagger)\supset D(\Hphys^\dagger)$ 
from $D(A)\subset D(\Hphys)$. 
Therefore $\Hphys^\dagger$ applies as 
$\Hphys^\dagger\Phi=\sx\Phi''$, 
where the second differentiation is in the sense of local absolute
continuity. 
We also have $D(\Hphys^\dagger)=D(\Hphys)$ 
because $A$ is Hermitian on $D(\Hphys)$. 

As a result, $\Hphys$ is {\sa} on its domain \Ref{eq-domain}.


\end{document}